\documentclass[conference]{IEEEtran}
\usepackage[utf8]{inputenc}

\usepackage{url}
\usepackage{cite}
\usepackage[flushleft]{threeparttable}
\usepackage[usenames,dvipsnames]{xcolor}
\usepackage{hyperref}
\usepackage{graphicx}
\usepackage{tabularx}
\usepackage{booktabs}
\newcommand{\tabitem}{~~\llap{\textbullet}~~}
\usepackage{microtype} 
\usepackage{ccicons}  

\ifCLASSINFOpdf
\else
\fi
\RequirePackage[normalem]{ulem} 
\RequirePackage{color}\definecolor{RED}{rgb}{1,0,0}\definecolor{BLUE}{rgb}{0,0,1} 
\providecommand{\DIFaddbegin}{} 
\providecommand{\DIFaddend}{} 
\providecommand{\DIFdelbegin}{} 
\providecommand{\DIFdelend}{} 
\providecommand{\DIFaddbeginFL}{} 
\providecommand{\DIFaddendFL}{} 
\providecommand{\DIFdelbeginFL}{} 
\providecommand{\DIFdelendFL}{} 
\newcommand{\DIFscaledelfig}{0.5}
\RequirePackage{settobox} 
\RequirePackage{letltxmacro} 
\newsavebox{\DIFdelgraphicsbox} 
\newlength{\DIFdelgraphicswidth} 
\newlength{\DIFdelgraphicsheight} 
\LetLtxMacro{\DIFOincludegraphics}{\includegraphics} 
\newcommand{\DIFaddincludegraphics}[2][]{{\color{blue}\fbox{\DIFOincludegraphics[#1]{#2}}}} 
\newcommand{\DIFdelincludegraphics}[2][]{
\sbox{\DIFdelgraphicsbox}{\DIFOincludegraphics[#1]{#2}}
\settoboxwidth{\DIFdelgraphicswidth}{\DIFdelgraphicsbox} 
\settoboxtotalheight{\DIFdelgraphicsheight}{\DIFdelgraphicsbox} 
\scalebox{\DIFscaledelfig}{
\parbox[b]{\DIFdelgraphicswidth}{\usebox{\DIFdelgraphicsbox}\\[-\baselineskip] \rule{\DIFdelgraphicswidth}{0em}}\llap{\resizebox{\DIFdelgraphicswidth}{\DIFdelgraphicsheight}{
\setlength{\unitlength}{\DIFdelgraphicswidth}
\begin{picture}(1,1)
\thicklines\linethickness{2pt} 
{\color[rgb]{1,0,0}\put(0,0){\framebox(1,1){}}}
{\color[rgb]{1,0,0}\put(0,0){\line( 1,1){1}}}
{\color[rgb]{1,0,0}\put(0,1){\line(1,-1){1}}}
\end{picture}
}\hspace*{3pt}}} 
} 
\LetLtxMacro{\DIFOaddbegin}{\DIFaddbegin} 
\LetLtxMacro{\DIFOaddend}{\DIFaddend} 
\LetLtxMacro{\DIFOdelbegin}{\DIFdelbegin} 
\LetLtxMacro{\DIFOdelend}{\DIFdelend} 
\DeclareRobustCommand{\DIFaddbegin}{\DIFOaddbegin \let\includegraphics\DIFaddincludegraphics} 
\DeclareRobustCommand{\DIFaddend}{\DIFOaddend \let\includegraphics\DIFOincludegraphics} 
\DeclareRobustCommand{\DIFdelbegin}{\DIFOdelbegin \let\includegraphics\DIFdelincludegraphics} 
\DeclareRobustCommand{\DIFdelend}{\DIFOaddend \let\includegraphics\DIFOincludegraphics} 
\LetLtxMacro{\DIFOaddbeginFL}{\DIFaddbeginFL} 
\LetLtxMacro{\DIFOaddendFL}{\DIFaddendFL} 
\LetLtxMacro{\DIFOdelbeginFL}{\DIFdelbeginFL} 
\LetLtxMacro{\DIFOdelendFL}{\DIFdelendFL} 
\DeclareRobustCommand{\DIFaddbeginFL}{\DIFOaddbeginFL \let\includegraphics\DIFaddincludegraphics} 
\DeclareRobustCommand{\DIFaddendFL}{\DIFOaddendFL \let\includegraphics\DIFOincludegraphics} 
\DeclareRobustCommand{\DIFdelbeginFL}{\DIFOdelbeginFL \let\includegraphics\DIFdelincludegraphics} 
\DeclareRobustCommand{\DIFdelendFL}{\DIFOaddendFL \let\includegraphics\DIFOincludegraphics} 
\RequirePackage{listings} 
\RequirePackage{color} 
\lstdefinelanguage{DIFcode}{ 
  moredelim=[il][\color{red}\sout]{\%DIF\ <\ }, 
  moredelim=[il][\color{blue}\uwave]{\%DIF\ >\ } 
} 
\lstdefinestyle{DIFverbatimstyle}{ 
	language=DIFcode, 
	basicstyle=\ttfamily, 
	columns=fullflexible, 
	keepspaces=true 
} 
\lstnewenvironment{DIFverbatim}{\lstset{style=DIFverbatimstyle}}{} 
\lstnewenvironment{DIFverbatim*}{\lstset{style=DIFverbatimstyle,showspaces=true}}{} 

\begin{document}
\bstctlcite{IEEEexample:BSTcontrol}
%
\title{``Should I Worry?" A Cross-Cultural Examination of Account Security Incident Response}

\author{Elissa M. Redmiles\\University of Maryland\\
eredmiles@cs.umd.edu}

%


\maketitle

\begin{abstract}
Digital security technology is able to identify and prevent many threats to users accounts. However, some threats remain that, to provide reliable security, require human intervention: e.g., through users paying attention to warning messages or completing secondary authentication procedures. While prior work has broadly explored people's mental models of digital security threats, we know little about users' precise, in-the-moment response process to in-the-wild threats. In this work, we conduct a series of qualitative interviews (n=67) with users who had recently experienced suspicious login incidents on their real Facebook accounts in order to explore this process of account security incident response. We find a common process across participants from five countries -- with differing online and offline cultures -- allowing us to identify areas for future technical development to best support user security. We provide additional insights on the unique nature of incident-response information seeking, known attacker threat models, and lessons learned from a large, cross-cultural qualitative study of digital security.
\end{abstract}


%
\IEEEpeerreviewmaketitle

\section{Introduction}
\label{sec:intro}
State of the art security technology can keep out many malicious actors from user accounts. However, automated detection and protection systems are not perfect; humans must still practice security behaviors: e.g., avoiding clicking through SSL warnings, completing secondary authentication tasks. 
Prior work has explored how best to design warning messages and phishing trainings to keep users away from dangers~\cite{egelman2008you,anderson2014users,akhawe2013alice,kauer2012not,Sunshine:2009:CWE:1855768.1855793,Bravo-Lillo:2013:YAP:2501604.2501610,wash2018provides,Kumaraguru:2007:PPP:1240624.1240760,Arachchilage:2013:GDF:2444047.2444325,Kumaraguru:2010:TJF:1754393.1754396,Sheng:2007:APD:1280680.1280692}. Other work has examined how people learn security behaviors from stories of negative experiences~\cite{rader2012stories,vaniea2014betrayed,redmiles2016think}, and yet other work has theoretically explored users' mental models of warnings or security~\cite{bravo2011bridging,asgharpour2007mental,kang2015my,wash2010folk,friedman2002users}. In these works, researchers typically simulate an attack in the lab, ask people to consider a theoretical scenario, or ask them to reflect on experiences that may have occurred in the distant past. While useful, theoretical approaches may lack ecological validity and requiring users to recall an incident long after it occurred may lead to omissions or distortions of collected data. As such, we have little understanding of how users respond when their real, valued accounts are threatened~\cite{wash2011influencing}. 

Our work takes a first step toward filling this gap: we conduct a series of open-ended qualitative interviews (n=67) soon after an in the wild security incident took place on participants' real Facebook accounts.\footnote{Facebook has over 2 billion users worldwide~\cite{constine2017facebook}, approximately two thirds of the estimated population currently online~\cite{international2017measuring}.} We specifically examine incidents in which Facebook identified and blocked a suspicious login to a user account, notified the user of this incident, and then asked the user to perform a secondary authentication task to regain access to their account. We use anonymized Facebook log records to identify and recruit participants, eliminating the need to recruit people based upon potentially biased self-reports. In our interviews, we asked participants to walk us through their experience of the incident. This included asking them about their feelings, information seeking processes, concerns over their account's compromise, and actions following the incident (which we validated, when possible, with log records).

To both facilitate broader generalizability and explore how incident response may be influenced by user and environment characteristics, we recruited participants from five countries that differ in online and offline cultural context: Brazil, Germany, India, the United States, and Vietnam. Based on rigorous analysis of the interview transcripts and timelines (Kripendorff's alpha = 0.87) we identify and instantiate a common process of account security incident response (Figure~\ref{fig:asp:process}). This process consists of incident awareness, mental model generation, and behavioral response. Variations in process execution are driven by the information users obtain through in-the-moment information seeking, users' threat models and past experiences, and three attributes of their cultural context: degree of Internet censorship, cultural collectivism -- as measured by Hofstede's indices~\cite{hofstede1983national} -- and differentiated platform use. 

We find that mental models are driven by the notification process in combination with participants? general threat models (what and who they fear online). While these mental models typically fall into either the category of false positives (e.g., this notification was caused by my actions, like logging in from a new location) and true positives (someone is attacking me), some participants formed surprising mental models, like perceiving the notification as a punitive action to correct misbehavior. Further, for participants who believed the incident about which they were notified was a real threat, wide variance in general threat models, emotional reactions to the notification and thought of attack, and differences in cultural and country-based Internet contexts resulted in a wide variance of protective behaviors. 

In addition to defining this process of account security incident response, we also qualitatively identify what made the suspicious login notification process we studied relatively effective\footnote{Over one third of participants in our study took what they perceived as a protective behavior after regaining access to their account (e.g., changing password, being more vigilant online).}. Chiefly, attention capture through a unique, interactive task and the creation of a sense of partnership between the user and the platform  We also identified areas for improvement such as leveraging users? own intuition and addressing known attackers when creating secondary authentication mechanisms.

\section{Related Work}
\label{sec:related}
Prior work on security incidents has focused heavily on the efficacy of warnings~\cite{akhawe2013alice,felt2015improving,egelman2013importance,sunshine2009crying}. Researchers have considered users' general mental models of the Internet~\cite{brandt2003insight,thatcher1998mental} and how those mental models influence security~\cite{kang2015my}. Further work has explored mental models of  security threats~\cite{asgharpour2007mental,camp2009mental,dourish2003security}; and Rader and Wash define a taxonomy of these threat models, which are unknown to the victim. Additional prior work provides a more general examination of user's negative experiences, including reflections on negative experiences with software updates~\cite{vaniea2014betrayed} and the effect of stories about negative security experiences on user behavior~\cite{Rader:2012:SIL:2335356.2335364}. 
We add to the body of knowledge on threat models by (a) defining the process by which threat models influence incident response, (b) illustrating how these threat models interact with information seeking behavior and emotional response, and (c) expanding the taxonomy of threat models by identifying four new folk models for known attackers (Section~\ref{sec:results}).

\textbf{Incident Response.} Most related to our work, Shay et al. surveyed U.S. Amazon Mechanical Turk crowd workers about whether they had ever experienced an account hijacking (someone taking over an email or social network account), to recall their concerns were about the incident, and to recall what they had thought had happened~\cite{shay2014my}. We build on their initial exploration by deeply examining how a multi-cultural population responded to a specific, in-the-wild incident, immediately after that incident; construct an end-to-end model of the process of incident response; and illustrate that incident response is not only influenced by the incident notification -- a factor also explored in the work of Shay et al. -- but is also heavily influenced by additional factors we identify, such as in-the-moment information seeking, past experiences, cultural context, and strength of mental models. 

Echoing the findings of Shay et al., we find that people have emotional reactions to security incidents, including fear and feelings of safety, that influence the process of incident response. Regarding protective behaviors, Shay et al. found that their users focused on passwords and password strength when forming mental models of how their account was compromised . In contrast, we find that while our participants mention passwords, they focus on a wide array of methods through which their accounts could have compromised and and also use a wide array of post-incident protective behaviors. These differences may in part be due to the difference in time between our two studies (2013 to 2018) or due to the differences in the demographics of our samples as U.S. users may have been trained to focus on passwords more so than users in other countries.

Additionally, both Shay et al. and Rainie et al.~\cite{rainie2013anonymity} found that U.S. users are primarily concerned about unknown attackers (e.g., hackers); while work by Kang et al.~\cite{kang2014privacy} on the privacy beliefs and views of Indian users showed an emphasis on government actors. The findings of our study echo Kang et al.'s findings that Indian users tend to focus on government attackers. We expand this out to find that a more general trend that this may be due to levels of internet censorship, as the Vietnamese participants in our study (also a censored country) also focused on government actors. Further, users in different countries may conceptualize their attackers differently: Rainie et al. break down attackers that the user knows based on their relationship to the user (friend, colleague), while our participants -- primarily those from non-western countries -- categorized their attackers by motivation (humiliation, spying for others). In combination, our work shows that users' mental models of attackers are quite complex: comprising the users' relationship to the attacker, the attackers' motivation, and what the attacker might access.
%

\textbf{Cross-cultural security.} Prior cross-cultural work in security has focused on comparisons between two countries (often the US vs. another country). Topics covered include enterprise security~\cite{chen2008cross}, Internet buying behavior~\cite{park2003cross}, online banking security~\cite{lim2010online,brown2005cross}, privacy beliefs~\cite{kang2014privacy}, and, in one case, general mental models~\cite{kumaraguru2005privacy}. A smaller set of papers has examined aspects of security across larger number of countries~\cite{harbach2016keep,chaudhary2015cross,de2016expert,karvonen2000cultures,sawaya2017self}. For example, Sawaya et al. surveyed users from seven different countries (not including South America or Southeast Asia) regarding their intention to behave securely, finding variance in intended action and security knowledge between cultures~\cite{sawaya2017self}. 
These works have all supported the need for a more nuanced examination of digital security from the lens of different user populations; a call echoed specifically for cross-cultural examinations in Crossler et al.'s position paper on future directions for behavioral cybersecurity research~\cite{crossler2013future}. 

Our work takes another step toward answering that call: here, we report the first interview study on digital security conducted across more than two countries. Considering multiple countries allowed for an important addition to cross-cultural security analysis: the consideration of culture as a set of dimensions (e.g., Internet penetration, Internet censorship, platform use types). These dimensions allowed us to identify potential effects of culture on user security more precisely than if we had considered only nationality as a singular representation of culture. For example, we were able to hypothesize about {\it why} we observe variations by nationality -- a gap in prior cross-cultural security work~\cite{sawaya2017self,de2016expert} -- and supports more concrete avenues for future work: e.g., developing tools that specifically protect users in censored countries from ``Digital Graffiti Artists'', by detecting a negative-sentiment post about a government leader and requesting SMS-confirmation before posting. 

We are not the first to propose consideration of cultural dimensions~\cite{hofstede1983national}. Traditionally, dimensions such as collectivism or masculinity were used to fully represent a particular culture, and this approach has been reasonably, questioned~\cite{mcsweeney2002hofstede}. We agree that dimensions cannot offer a full picture of culture. However, we do suggest that consideration of domain-specific dimensions (e.g., censorship, penetration) can be a valuable {\it part} of cross-cultural security work. As such, developing a taxonomy of cultural dimensions relevant to security may be a fruitful direction for future work.

Moving forward, our results suggest an increased need for cross-cultural work. As aforementioned, combining results of prior, U.S. studies of incident response with our own cross-cultural work allowed us to enrich our understanding of user threats and discover new, relevant security factors. That said, we acknowledge that cross-cultural work can be quite difficult or, at times, infeasible due to monetary limitations or other restrictions. In these cases, we encourage appropriate description of the generalizability of the results and support for cultural-expansion studies that replicate existing study designs on participants from different cultural or geographic contexts. 

\section{Methodology}
\label{sec:methods}
To investigate the process of in the wild security incident response, we conducted 15 pilot interviews and 67 non-pilot interviews with Facebook users from Brazil (BR), Germany (DE), India (IN), the United States (US), and Vietnam (VN) who had been notified of a real suspicious login incident on their Facebook account during August 2017. Our research procedures were approved through our ethics review process. 

Here, we describe our sampling and recruitment methodology, interview process, data analysis and validation procedure, and the limitations of our work. 

\subsection{Sampling and Recruitment}
\begin{figure}
\small
\centering
\includegraphics[width=0.48\columnwidth]{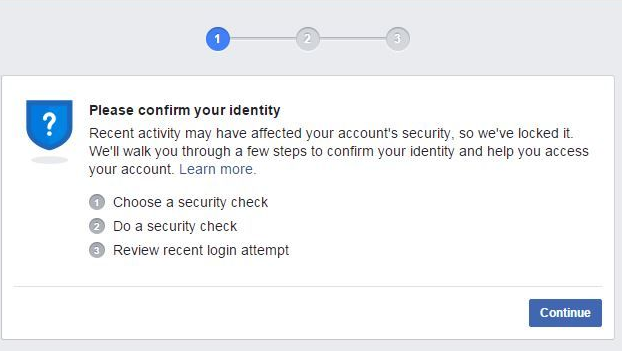}\hfill
\includegraphics[width=0.4\columnwidth]{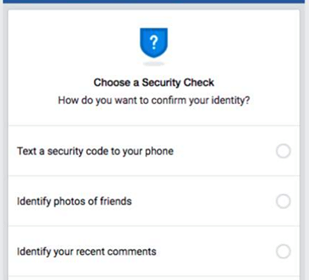}
\vspace{-2ex}
\caption{\label{fig:suspicious} (Left) The screen notifying users that a suspicious login has occurred. When available this screen includes information about the geographic location of the login. (Right) The screen in which participants can chose which secondary authentication task to complete.}
\vspace{-1ex}
\end{figure}

To maximize ecological validity and minimize self-report biases that may be introduced if we were to identify eligible participants by asking people to report whether they had experienced a particular incident~\footnote{Prior work shows that people with different socioeconomic backgrounds report security and privacy incidents at very different rates~\cite{redmiles2017digital}.} we consulted anonymized Facebook log records to identify people whose accounts had a suspicious login incident. To ensure comparability of data across participants and countries we sought to recruit people who had experienced the same type of account security incident during the same time period (two weeks). Suspicious login incidents are identified by Facebook machine learning systems that monitor for deviations from typical login patterns. Specifically, we selected only those accounts for which the classifier was most confident that the suspicious login incident was authentic (not caused by some non-malicious owner action). Facebook users who have a suspicious login (which the classifier may have predicted with variable confidence) to their account are blocked from logging in (or continuing account use) with a message that notifies them of the suspicious login. The user must then immediately complete a secondary authentication process (Figure~\ref{fig:suspicious}) to regain access to their account. Secondary authentication requires users to complete one of a set of possible tasks, for example, identifying pictures of their friends, identifying people with whom they have recently messaged on Facebook messenger, or using two-factor authentication if it is enabled. 

\begin{figure}
\centering
\includegraphics[width=0.9\columnwidth]{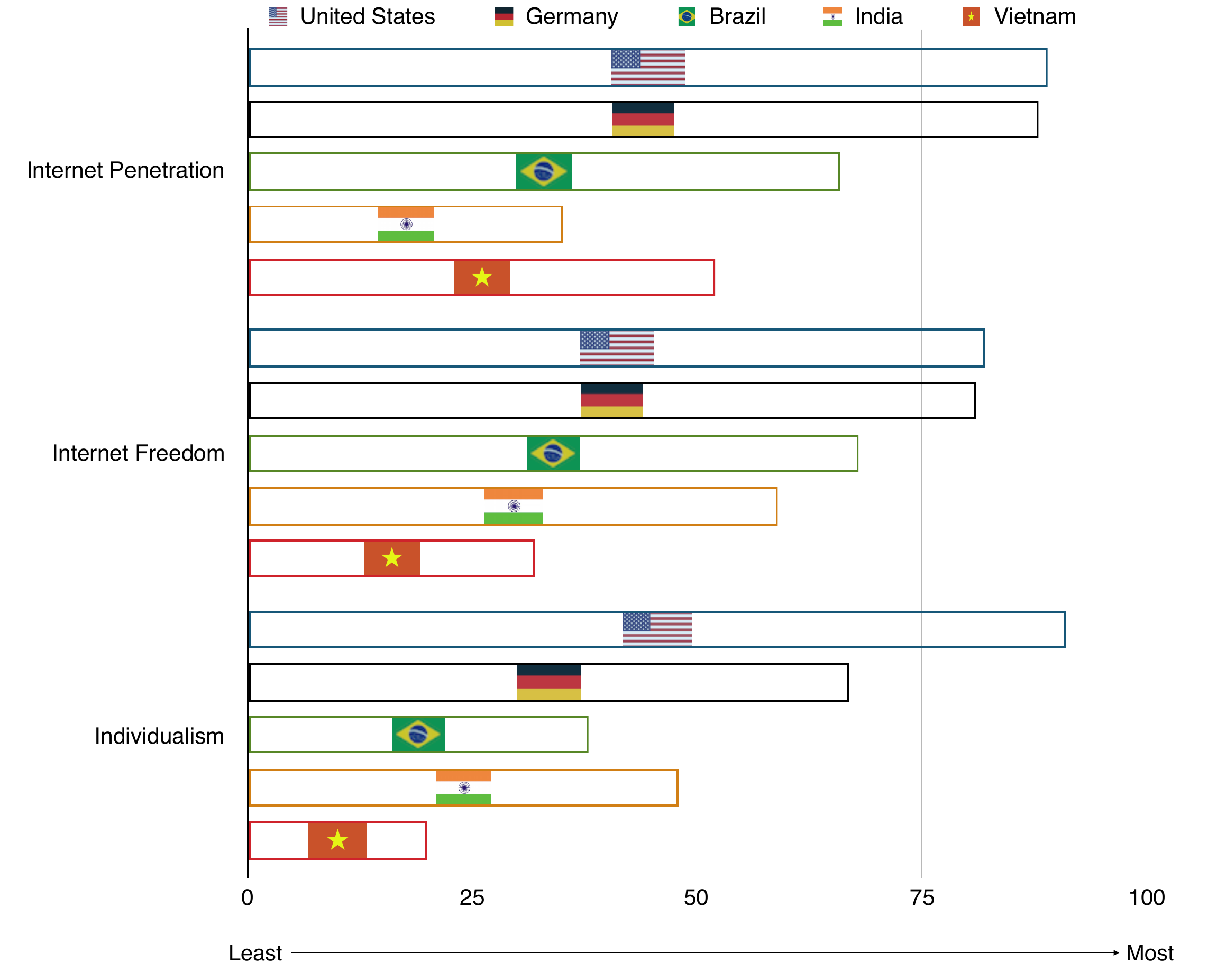}
\vspace{-2ex}
\caption{\label{fig:countryattributes} Internet penetration~\cite{Internet2016Internet}, Internet freedom~\cite{house2017freedom}, and Individualism~\cite{insights2017country,hofstede2011dimensionalizing}, for each of the five countries from which we recruited interview participants. Metrics are normalized such that a lower score indicates less penetration, less freedom, and less emphasis on the individual (collectivism).}
\vspace{-2ex}
\end{figure}

We specifically sought to study participants from five different countries, each with a different cultural and technological profile, to maximize generalizability. We selected Brazil, Germany, India, the US, and Vietnam as these countries span four different continents and differ in Internet penetration, Internet freedom, Facebook adoption level and incidence of suspicious account logins\footnote{We are not able to disclose these adoption and incidence figures.}, as well as cultural characteristics such as emphasis on the individual vs. the collective (e.g., collectivism), as shown in Figure~\ref{fig:countryattributes}. 

Of our eligible participant sample, we identified those whose account geolocation data indicated that they were within 100 miles of one of our interview locations, who had an email address registered on Facebook, and who had not previously opted out of being contacted about research via this email address. We selected two interview locations per country: one in a major city (e.g., New York City) and one in a small town one to two hours from that city, in order to ensure a diversity of participants. As we sought to recruit 75 participants in five countries (Brazil, Germany, India, Vietnam, and the US) and time zones, we made eligible for an interview anyone who had experienced an account security incident in the two weeks prior to the first interview day for their country. There were two consecutive days of interviews for each country; all interviews were conducted within a two week period.

We contacted eligible participants via the email associated with their account. Recruitment emails were sent from Facebook email addresses using a Facebook letterhead to indicate authenticity; all emails were translated into the participants' locale (location-based language). The email explained how their contact information was obtained, that a team of researchers was conducting a study of people's experiences on Facebook, and that they were eligible for the study if they were interested in participating. The email noted the study method (in-person interviews), length (30 minutes), and compensation (\$75~\footnote{Compensation for interviews conducted by UX consultancy firms such as those used to schedule and interview participants in this study is typically higher than for academic interviews, especially in studies such as ours where study requirements make scheduling very tight -- e.g., we wanted to conduct 15 interviews within two synchronous days.}  or equivalent). Those who were interested in participating completed a short demographic and scheduling survey.\footnote{Scheduled participants completed a consent form prior to attending their interview. There was no penalty for opting out, and record of their choice to opt in or out was not associated with their Facebook account unless they requested not to ever be contacted for Facebook studies.} We selected participants in an attempt to maximize diversity in terms of gender (M/F), age, and educational attainment. 

\subsection{Interview Protocol} 
At the beginning of the interview participants were reassured that there were no right or wrong answers and that the moderator was just interested in their opinions. Participants were asked about what they typically do on Facebook. Then, they were asked whether they remembered anything happening recently that was out of the ordinary when they were logging into their account or trying to use it. If they did not immediately recall the incident the interviewer asked if they happened to have done something like the secondary login process that we knew they had completed. After this prompting, all participants remembered the incident.

\textbf{Incident Walkthrough.} The participant was then asked to tell the moderator about what had happened. The moderator then paused, and explained to the participant that the moderator was going to draw a  timeline on a piece of paper, to make sure they captured what the participant was saying. The moderator drew a line and placed the notification in the middle, they then added any events already described by the participant (e.g., behaviors afterwards, feelings at the time) and confirmed that these were accurate with the participant. See Figure \ref{fig:br4} for an example. 

Moderators then queried respondents':
\begin{itemize}
\item feelings: about the experience (at the time of the notification and at the time of the interview)
\item mental models: what they think happened and why, and how they figured it out
\item incident response: what they did during and after seeing the notification, including security behaviors (e.g., changing password) and information seeking
\end{itemize}
Interview questions were asked in different order depending on the way that participants initially described their experience, in order to create as natural and conversational a flow as possible. Relevant information from participant's answers was iteratively added to the incident timeline. 

\textbf{Threat Model Assessment.} Finally, if not already mentioned, the moderator asked participants to detail who they would be most worried about gaining access to their account (e.g., friend, stranger) and why (e.g., what they would be concerned about this person doing or accessing).

\subsection{Interview Process}
The interviews were conducted in the country's official language. That is, US interviews were conducted in English, German interviews in German, Indian interviews in Hindi, Vietnam interviews in Vietnamese, and Brazilian interviews in Portuguese. To enhance consistency, all interviews, including those conducted in the US were completed by highly trained moderators who spoke both English and the native language of the country fluently.

\textbf{Moderator Training and Interview Consistency.} After developing the interview protocol, the researchers tested the interview protocol through 10 pilot interviews with U.S. participants and iteratively improved the protocol until it was consistently understood by participants and new issues stopped emerging. The researchers then distributed the final interview protocol to the moderators in each country as well as to an interview training manager who had extensive experience ensuring consistency across multi-country, multi-language interview studies. All moderators had at least three years of experience conducting UX research interviews in the country native language. To ensure that the protocol would be implemented as consistently and accurately as possible between moderators, a three-step, moderator training and validation process was conducted. 

First, the researchers met with the interview training manager and talked through the goals for the protocol. 

Once the researchers were satisfied that the interview training manager fully understood the protocol and would conduct interviews in a comparable manner, the interview training manager then repeated this process with each of the country interview moderators. The interview manager also conducted a test interview with each moderator in the role of interviewer, and the manager in the role of participant. After the interview training manager was satisfied with the moderators ability to comparably conduct interviews according to the protocol, the researchers met with each moderator and reviewed their understanding of the interview protocol and method of asking questions, providing additional feedback until all moderators were well prepared. 

Finally, the first interview of the first day of each of the country interview sessions was treated as a pilot interview. This pilot interview was simultaneously translated. Both the researchers and the interview training manager listened to the interviews and provided feedback to the moderator immediately following the session and before the start of the non-pilot interviews that were ultimately analyzed. All pilot interviews met our standards for consistency and quality, and thus we proceeded with the non-pilot interviews, as planned. 

\subsection{Data Analysis and Validation}
\label{sec:method:analysis}
Interview recordings were transcribed and translated into English. We then analyzed the interview data, including the timelines, via a qualitative open-coding process~\cite{strauss1990basics}. Two researchers first reviewed seven of the transcripts and iteratively generated a codebook. They then independently coded the remaining 60 interview transcripts and timelines; and compared their codes using Krippendorff's alpha, the recommended metric for checking the validity of qualitative interview coding. The Krippendorff's alpha of this study is 0.87, which is above the required threshold of validity~\cite{lombard2002content}. Further, after calculating the agreement metric, the interviewers reached 100\% agreement on the final codes for each interview. In the results, we state the number of participants who expressed each theme, rather than using percentage values, to avoid over-implication of generalizability. 

\textbf{Data Validation.} During the interviews, participants often attributed receiving the account security notification to a particular action taken by themselves (e.g., logging in from a new device) or by someone else (e.g., an attempted ``hack''). Unfortunately, while Facebook classifiers can predict a potential unauthorized access with existing technology it is difficult to validate the accuracy of those predictions, aside from using user self reports. As such, we cannot use log data to validate whether participants' mental models (including their timelines reflecting back on what may have caused the incident) were accurate. However, we were able to use internal log records to validate the completion of some behaviors users reported doing after the incident. Specifically, we could validate whether they actually changed their password or their privacy / security settings. We found that, of the 14 participants who reported doing so, 12 actually had done so, and one of the two who had not had done so four weeks prior to the incident, and thus may have experienced telescoping bias, in which participants may perceive older events as being more recent than they really are~\cite{sudman1973effects}. Our finding: that 12 of 14 reports were perfectly accurate offers support for the credibility of our data, and evidence in confirmation of the validity of security interview data, at least when collected soon after a security incident.

\subsection{Sample Descriptives}
We had 15 non-pilot participants in Brazil, 11 in Germany, 15 in India, 17 in Vietnam, and 9 in the US. In all countries we aimed to recruit 15 participants, in order to achieve sample sizes within best-practice recommendations for qualitative research. Table~\ref{tab:demo} shows the demographics of our participants. 
%

\begin{table}
\centering
\small
\caption{\label{tab:demo} Participant Demographics}
\begin{threeparttable}
\begin{tabular}{lccccc}
& Gender & Education\tnote{1} & Age\tnote{2} & FB Use\tnote{3}\\
\toprule
BR & 8M/7F & 9PS/6HS & 10M/5X/0B & 14V/10F/7M/3B\\
DE & 6M/5F & 5PS/6HS & 3M/7X/1B & 9V/12F/4M/3B\\
IN & 14M/1F & 11PS/4HS&  14M/1X/0B& 9V/8F/4M/3B\\
US & 4M/5F & 3PS/6HS & 4M/4X/1B& 13V/5F/9M/1B\\
VN & 12M/5F & 7PS/10HS & 14M/3X/0B& 12V/8F/8M/7B\\
\bottomrule
\end{tabular}
\begin{tablenotes}
     \small
      \item[1] PS: post-secondary education (some college or above), HS: high school diploma or less education
      \item[2] M: post-Millennial and Millennial~\cite{dimock2018defining} (ages 18-37), X: Gen X (ages 38-54), and B: Baby Boomer (ages 55+)
      \item[3] Using Facebook to V: view content (passive), F: connect with friends, M: Facebook Messenger, and B: run a business.
    \end{tablenotes}
   \end{threeparttable}
   \vspace{-0.4cm}
\end{table}

\subsection{Limitations}
\label{sec:methods:limit}
While we attempted to obtain a sample in each country that was demographically diverse in terms of gender, education, and age, we were unable to do so in India and Vietnam. We hypothesize that this is due to cultural norms around gender in both countries. In the future, we would recommend that in-person interview studies seeking to recruit women in places with these norms should consider in-house interviews, rather than office-based interviews. Further, researchers should remain open to a potential need for snowball sampling to obtain an appropriate sample. 

We also did not achieve our goal sample size in Germany and the U.S., partially due to weather issues in the US (heavy storms) and lower incident incidence rates. We did however reach theme saturation in each country, typically after seven interviews. Further, prior work has most thoroughly studied security in Western countries, as such, we acknowledge this limitation in our work, but believe it does not undermine the final, qualitative results.

Additionally, we examined a single type of account security incident that occurred naturally on a single platform: Facebook. While Facebook is heavily adopted, we cannot necessarily generalize our findings to all other types of accounts or incidents (e.g., phishing). We encourage future work evaluating and expanding the process model we define to include other incidents and platforms. Like most self-report studies, we cannot validate the accuracy of participant's responses. However, we know for certain from log data that our participants all experienced a notification about an account security incident. Additionally, we validate the behavioral reports, where possible, a first to our knowledge in interview studies; the high validity level of the reports (Section~\ref{sec:method:analysis}) supports the integrity of our interview data. Finally, participants knew that this research was being conducted by Facebook, and this knowledge may have biased their answers. To mitigate this effect, none of the interviews were conducted by Facebook employees, and, at the beginning of the interview, participants were told that the interviewer was not from Facebook and that the interviewer wanted to hear anything -- good or bad -- about the participants' experiences. Further, the participant was reassured that their answers would never be associated with their Facebook account and would have no effect on their Facebook account.

\section{Account Security Incident Response Process}
\label{sec:results}
\begin{figure*}[h!]
\small
\centering
\includegraphics[width=\textwidth]{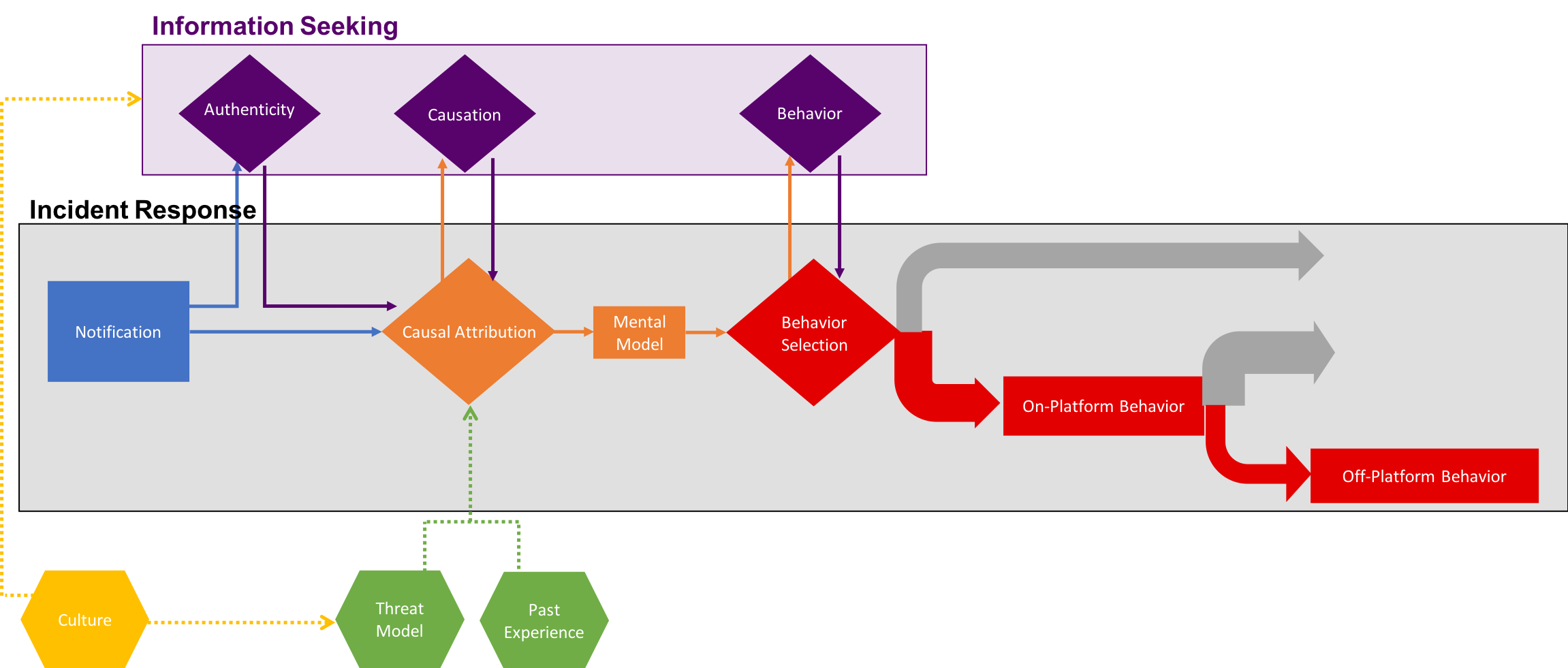}
\caption{\label{fig:asp:process} Process diagram illustrating the generalized process of account security incident response we observed across our 67 participants. The rectangles represent inputs or outputs (e.g., the notification message, the cause of the incident, a particular behavioral response), the diamonds represent a user-driven process (e.g., decision-making, information seeking), and the hexagons represent influencing factors (e.g., past experiences).}
\vspace{-1ex}
\end{figure*}
Our interview analysis revealed a common process of account security incident response across our 67 participants. The process consists of three main steps: becoming aware of the incident, causal attribution (developing a mental model of the incident), and behavioral response (e.g., changing their password on Facebook, changing their password on other websites). The decisions users make in each step of the process are influenced by external information -- that they may seek out at each step -- as well as their past experiences, threat models, and cultural contexts. In this section we describe this process, summarized in Figure~\ref{fig:asp:process}. In Section~\ref{sec:info} we explore how the information-seeking practices we observe differ from general security advice-seeking practices studied in prior work; and in Section~\ref{sec:culture} we discuss the influence of cultural context on both the response process and the information seeking sub-process.

\subsection{Incident Awareness}
\label{sec:results:notification}
The incident response process begins with the user gaining awareness that an incident has occurred. This is done through a message (Figure~\ref{fig:suspicious}) informing the user that a suspicious login has occurred. After seeing this message, users are required to complete a secondary login process, which users must complete to regain access to their Facebook account. Participants were offered a choice of secondary authentication options, such as identifying pictures of their friends, identifying people they had recently messaged, or two-factor authentication (Figure~\ref{fig:suspicious}).

\textbf{Awareness is triggered by the unique authentication process more often than by the notification message itself.} The majority of participants (48 of 67) did not describe becoming aware of the incident from the notification message, but rather described that their awareness of the incident was triggered when confronted with the secondary authentication task. For example, VN12 describes how they learned about the recent incident with no mention of the initial notification message: ``it asked me to log in again and asked me to identify friends...so I knew something was wrong, it had said something about suspicious login to my account.'' As exemplified by this quote, once attention was captured participants then reflected on what the notification message had actually said.

We hypothesize that users may be so habituated to all types of ``warning'' messages, as has been shown in a plethora of other work on SSL warnings~\cite{egelman2008you,anderson2014users}, that they simply click through the message and their attention is only captured when they must perform a task. Our findings parallel findings that SSL warnings were more effective (able to combat habituation) when people had to take an action (e.g., highlight text) in order to proceed. This suggests that, when merited, incident secondary authentication tasks as part of the notification process may be effective and beneficial for both users and platforms; perhaps more so than warning messages or more stringent security requirements, both of which may be met with user resistance and negative platform sentiment~\cite{vaniea2014betrayed}. That said, platforms must be careful of over-acting ``security theater''~\cite{schneier2007praise} and causing users to feel a false sense of safety, which may reduce their likelihood of protecting themselves when necessary.

The remaining minority of participants became alert to the potential incident from the notification message itself. For example, BR14 noticed the message and the information described peaked their attention, ``Facebook said that there was someone in Brasilia [a different city in Brazil] trying to access my account, and I was at home, so [I thought]...that can't be right.'' Throughout the remainder of the paper we use the term ``notification'' to refer to whichever portion of the notification flow participants referenced as what alerted them that there was potentially something wrong with their account.

\textbf{Respondents felt surprised, fearful, and/or annoyed.} Some participants were surprised (8 of 67) when alerted to the incident. They expressed a lack of awareness that Facebook conducted these types of ``checks'': US1 says, ``I was surprised at first just because I had never seen that specific type of check before. Maybe I see like emails or something but I had never seen that before from Facebook.'' 

17 participants expressed fear in response to the notification for one (or both) of two reasons. Eleven participants felt afraid because they feared that someone had gained access to the account. These participants went through the mental modeling process immediately upon gaining awareness and swiftly concluded that the incident was caused by an attacker. On the other hand, eight of the 17 were afraid they would not be able to complete the secondary authentication process and regain access to their account; these participants were concerned with {\it how} they would get back into their account before being concerned about {\it why} or {\it what} had happened. IN6 for example says ``I was afraid, what if I don't know the comments [that I have to identify], and I lose my account? That's the only place I can run my business''. Three of the eight participants who had this concern did, in fact, need to attempt more than one secondary login task before successfully completing one task and regaining access to their account.

Finally, annoyance was only observed among 14 participants who received the notifications frequently, the majority of whom concluded that the notifications were due to routine Internet use (Section~\ref{sec:results:mental:model} provides more detail on the mental models of these participants). Our findings echo prior work showing that repeated warnings elicit annoyance~\cite{akhawe2013alice, herley2009so}, while account hijacking incidents may elicit fear~\cite{shay2014my}.

\textbf{Secondary login process can create a sense of partnership.} Many participants felt positively about the fact that they were alerted to an incident. A total of 32 participants felt positively about the notification, eight participants also had other emotions such as fear, surprise or annoyance while the remainder felt only positively. These participants perceived the notification process as a form of active partnership with the platform: Facebook was watching out for threats to their account and they as the user were completing a task (the secondary authentication task) to help Facebook keep them secure. BR11 explains that now that she knew Facebook was ``on my side'' and that ``I don't think anyone else could do [the secondary authentication task], I felt safe after [that]''. Similarly, DE1 says, ``it made me feel like...[Facebook] is on top of the game...somebody is watching out to make sure I don't get hacked'' and DE3 says that now that he knows about the notification process and has done the secondary authentication task, ``I feel much safer about my account.'' The fact that participants felt that the secondary task they had to complete was secure - the vast majority (46 of 67) trusted the method of secondary authentication ``it seemed much safer than those security questions'' (US5) - appears to further enhance this sense of partnership. This finding expands on prior findings, which considered multiple platforms, and showed that password reuse notifications made users feel safe~\cite{golla2018site}.

\textbf{A few respondents sought out information about authenticity.} This sense of partnership clearly relies on participants believing that the notification and task were actually from Facebook. While the majority of respondents immediately trusted the notification and/or secondary authentication task, a few wondered -- is this notification malicious? -- that is, they sought out information about the {\it authenticity} of the notification. DE7 said ``At first, I thought, maybe this is someone trying to hack me, to find out about my friends. I've never seen this on Facebook before, maybe it got hacked. So then I looked it up and found a page from Facebook explaining what this is.'' Five of 67 participants (four of whom expressed surprise in reaction to the notification) 
sought out information about message authenticity; and three other participants raised authenticity concerns, but ``trusted that [the notification] was probably ok'' (IN6) and completed the secondary authentication process without seeking out additional information about authenticity. The fact that these concerns were raised by eight participants is, in fact, encouraging: users are thinking critically about what information is being asked of them. This also underscores the importance of having easily accessible information (e.g., the Facebook help page~\cite{fbsecuritycheck} that DE7 consulted) that people can use to verify notification authenticity.

\subsection{Mental Model Generation}
\label{sec:results:mental:model}
Participants try to develop a mental model -- a rough internal representation of how something happened or how something works -- of the incident either while being notified (after gaining awareness of the incident, but before regaining access to their account) or after they have regained access to their account, but before taking a protective behavior to prevent future incidents.

\textbf{Participants identify one of three possible causes for the incident: something they did, an attacker attempting to gain access, or ``random selection'' by Facebook.} Overall, 29 of our participants thought the incident about which they were being notified was caused by something they did (i.e., they thought the incident notification was a false positive), while 31 thought the incident was caused by an attacker (i.e., a true positive), and 7 did not associate the incident with anything about their account, but rather thought that the notification / secondary authentication task was either randomly assigned or assigned to all Facebook users (an inaccurate mental model). As discussed below, some participants reached multiple conclusions about the cause of the incident, within these broader categories (Table~\ref{tab:models}).
\begin{table}
 \centering
  \caption{\label{tab:models} Incident Mental Models}
 \begin{tabular}{ll}
  \toprule
  ``Something I did'' (False Positive)\\
  \tabitem Because I logged in from new location & 9 \\
   \tabitem Because I was doing something unsafe or not allowed & 9 \\
  \tabitem Because I used a new or rarely-used device & 7 \\
  \tabitem Because I mistyped my password & 4\\
  \tabitem Because I use a VPN or private browsing & 6\\
\midrule
  ``An Attack'' (True Positive)\\ 
   \tabitem By someone I don't know (Unknown Attacker) & 10 \\
   \tabitem By someone I know (Known Attacker) & 13\\
   \tabitem By either a known or unknown attacker & 8\\ 
\midrule
  ``Random Selection'' \\ 
   \tabitem Random security check (``like TSA'') & 4 \\
   \tabitem New security measure for everyone & 3\\
   \bottomrule
 \end{tabular}
 \vspace{-2ex}
\end{table}

Of those who thought that the incident was caused by something they did, they attributed the notification to logging in from a new geographic location, logging in on a new device (or from multiple devices), mistyping their password, or using VPN or private browsing mode. Specifically, nine participants attributed the suspicious login notification to logging in from a new location. For example, BR12 says ``Yeah, it's because...it seemed that I logged in from another country. So it appeared to Facebook that there was a suspicious access.'' Seven participants thought that the incident was caused by using a new device, or multiple devices. An additional four participants thought that the incident was caused by mistyping their password, sometimes repeatedly: ``well, I reset my password like four months ago and then I forgot that I did that, so I just retyped my old password like 10 times, and then it happened,'' reported IN9. Finally, six participants thought that the notification appeared because they sometimes used a VPN or private browsing mode on their browser to access Facebook. 

Finally, nine participants thought that the incident was caused by doing something ``bad'': doing something insecure or something of which the platform would disapprove. For example, DE1 says: ``I thought about it that maybe I did something wrong. Did I click on a link or did I get an email?'' VN1 attributed the account security incident to having done something that Facebook would not approve of the night before he received the notification: ``I hacked likes. So basically, I just hacked number of likes on the post,'' which VN1 explains means that they ran a scheme with their friends to all like each other's posts. He explained that because Facebook had figured out he was doing this, they probably made him do this extra login procedure. In reality, these nine users were incorrect, as, while Facebook does also have processes to protect users from spam and malicious actors, those processes would not have resulted in a suspicious login notification.

Seven participants had a variety of less accurate mental models, all of which illustrated a misperception of the incident. Four of these participants thought that the notification and secondary authentication task was ``a random security check, like TSA does at the airport'' (US2) or that Facebook was performing ``like a checkup to make sure [the] account was ok'' (BR7). The remaining three believed that the notification and secondary authentication task was given to all Facebook users, potentially to increase Facebook'a reputation or to counteract a current issue, like Fake News. For example, IN4 says ``I hear about fake news a lot...So, I just merely think that they are cracking down or something, and everyone had to do this extra thing, that's what I thought.'' 

\textbf{General threat models influence incident-specific understandings.} For those participants (31 of 67) who thought the incident was caused by someone trying to gain access to their account, their conception of what was happening drew heavily on their general threat model for their Facebook account. In general, participants threat models for their Facebook accounts consisted of (a) who they were worried about gaining access to the account and (b) what they were worried about an attacker accessing. Across all participants, 25 were concerned about unknown attackers, 19 about known attackers, and the remaining 23 about both. Of the 31 participants who thought this specific incident was caused by an attacker, 10 thought that an unknown attacker was attempting to gain access, 13 thought that it was a known attacker, and 8 mentioned both types of attackers -- each of these conceptions was in line with their general mental model. That is, if they were generally concerned about an unknown attacker and they thought that this incident was caused by an attacker, rather than e.g., something they did, they thought that the attacker was unknown to them. Similarly, those general Facebook threat models centered around someone they knew (a known attacker) gaining access to their account and who thought that this specific incident was caused by an attacker, attributed the incident to a particular person they knew: ``I have been going through a breakup and he's real savvy with devices...I think it was him'' (VN3). 

Participants whose threat models centered around unknown attackers --``someone bad trying to get in'' (US6) -- expressed conceptions of these attackers that fell broadly into the theoretical framework previously defined by Wash~\cite{wash2010folk}: unknown attackers were viewed as either ``digital graffiti artists'' -- who gain access to accounts in order to show off -- or ``burglars'' -- who gain access to accounts for the purpose of theft. On the other hand, those whose threat models centered around a known attacker had a broader set of mental models. While some conceived of their known attackers in the ways proposed by Wash's existing framework -- as ``digital graffiti artists'' and ``burglars'' -- others participants described a broader, new taxonomy of \textbf{known attacker folk models}:
\begin{itemize}
\item \textit{the Spy:} Some participants expressed concern that someone they knew would attempt to gain access to their account in order to provide information to the government about them (e.g., information obtained by reading their private messages, or information about their business). For example VN10 describes a conversation with a ``hacker'' friend who he knows provides information to the government: ``I mentioned about this situation [to my friend]...my friend said that I should talk to this other hacker and show this code and tell that I am this person, I'm his friend...and then this other hacker said he was sorry and stopped doing that.'' The legitimacy of this threat model is supported by reports of government attacks~\cite{spy:citizenlab}.
\item \textit{the Snoop:} Other participants expressed concern that someone they knew would want gossip or to know something private about them, not something that was financially valuable, but something personal (e.g., how a romantic relationship  was going). BR12 suspected her best friend of snooping, she says ``My best friend, why would she have done this? She should have asked me, I would have given her permission or told her.''
\item \textit{the Who Else:} In an variation on the ``Big Fish'' model~\cite{wash2010folk}, some participants explained that they could not understand {\it how} an unknown person could gain access to their account without knowing them -- nor why they would want access -- and therefore concluded that the attacker must be someone they know. IN3 explains: ``It is totally impossible that [an unknown person] gets my account, or he wants my account, he can't get into my account because he is totally unknown and he doesn't know anything about me.''
\item \textit{the Humiliator:} Finally, participants thought that someone they knew might want to access their account to humiliate them: ``if he just did it for fun, or to show me he could, that's ok. But if it's not just for fun, then I'm afraid that he would...upload photos or videos to humiliate me.'' (VN13) This is an expansion of the ``Digital Graffiti Artist'' model~\cite{wash2010folk}, where users are ok with someone they know doing some types of ``graffiti'' (e.g., unauthorized posting) but not others (e.g., humiliation).
\end{itemize}
Of the 19 participants who were concerned about known attackers, their conception of these known attackers was relatively equally distributed across this taxonomy of motivations, suggesting that participants are not merely trying to explain away notifications (e.g., the Who Else model) but are genuinely concerned about attacks from those they know. 

Finally, participants' mental models concern not only who is attacking them but nature of the attack: 46 participants were worried about someone acting as them on their profile, 26 were worried about someone accessing their messages, 23 were worried about someone getting hold of their PII or their pictures, respectively, and 10 were concerned about financial loss or being reported to the government, respectively. These concerns informed participants mental models of {\it why} the attacker was attempting to gain access and thus influenced whether participants chose to take protective action after regaining access to their accounts, and which actions they took, as described in Section~\ref{sec:results:behavior}.

\textbf{Past experiences with similar incidents may reduce perception of threat.} Participants who reported having seen a similar incident notification on Facebook in the past (14 participants) often reported a shift in their mental model of the incidents over time: while at first they thought that an attacker was attempting to gain access, after repeated checkpoints they started viewing the incidents as ``routine'' (DE5). DE2 explains, ``the first time, I was worried...[now I understand] Facebook asks all users this when they go into a foreign country [now] I don't think it has to do with me.'' Similarly, VN6 explains that she was originally concerned for her account, but subsequent checkpoints made her think that this was just a routine security check: ``The first time that it appeared, I thought it was someone who was trying to access to my Facebook but the next times, I realized that it was Facebook [trying] to enhance the security.'' Thus, multiple notifications are a signal to users that the system may be generating false positive reports. Some participants may indeed see the incident notifications frequently due to false positives -- e.g., caused by use of privacy enhancing technologies (VPN, private browsing)~\cite{davidson2018privacy} or frequent travel or multiple device use, for example, VN12 explains, ``Because I have like two accounts. I have never encountered such problem with one account, but with the other account, I always encounter that problem. Because I only use one account on laptop, so that thing has never happened. I have never encountered such request. But with the other account, I use it on various computers, so that?s why it always requests to verify if that was me.'' However, this does not mean that an authentic threat cannot also occur, even after prior false positives.

\textbf{Participants only mention past experiences with Facebook, not other platforms.} It is interesting to note that experiences on {\it other} platforms did not appear to influence mental models of the incidents we studied. In fact, no participants mentioned similar experiences on other platforms, they only referenced prior experiences on Facebook. This is perhaps surprising, as prior work has suggested that negative experiences, or stories about those experiences, can generally inform users' security posture. We hypothesize that this may have occurred for two reasons. First, our incidents are not precisely {\it negative experiences} but rather {\it prevented} negative experiences, which are perhaps less powerful. Second, most participants considered the secondary authentication process quite unique -- as DE1 explains, ``there is no other provider doing this type of authorization'' -- it is thus possible that the uniqueness of the secondary authentication task may have prevented generalization from other prior experiences. Our findings thus raise questions about what precise types of negative experiences generalize across platforms, and the level of similarity between experiences required to allow users to make connections, whether positive -- behavior transfer -- or negative -- fatigue transfer -- between them.

\textbf{Participants may reach out for support as part of the process of mental modeling.} In addition to drawing on their own understandings and past experiences to form their incident mental models, 17 participants sought out additional information to help them understand the incident. They sought information about {\it causation}: ``I wanted to know, why is this happening? What happened to my account?'' (VN2). The information they collected, either from other people or through online sources ({\it how} people sought out information is discussed in more detail in Section~\ref{sec:info}) influenced the cause they ultimately attributed to the incident and whether, and what, protective behavior they took. For example, US7 says, ``well, I searched on Google, and it said that sometimes there are these people online, and they just try getting into a bunch of accounts. And so I thought wow, that's probably what's happening here...At first I thought it was no big deal, but then after reading that, I thought, wow, I should probably do something about that.''

\textbf{Despite the majority reaching plausible mental models, many oscillated between multiple possible models.} Finally, the mental models that many participants developed through their process of causal reasoning were what we characterize as weak. That is, of the 51 participants who had plausible mental models of the incident (that it as caused by something they did or by an attacker), 27 offered up multiple possible models or hypothetical causes for the incident, or repeatedly caveated the mental model they described with ``I don't know'' or other statements of low confidence. For example, US4 provides a hypothesis for why they saw the notification but notes they are not too sure they are correct, ``I think, maybe they see I'm logging in two locations, I don't know, honestly, but I just go on with it.'' Similarly, some participants, like IN11, offered multiple possible explanations for what happened ``Well, my thought is that maybe I have accessed my Facebook account... either from an unsecured line. Or I might have shared certain things, which I shouldn't have done. Or I shared some certain details to my friends and colleagues and they try to check the account from a different location or something like that.'' As described further in the next section, these oscillations led to uncertainty about what to do next. 

\begin{figure}[t]
\small
\centering
\includegraphics[width=\columnwidth]{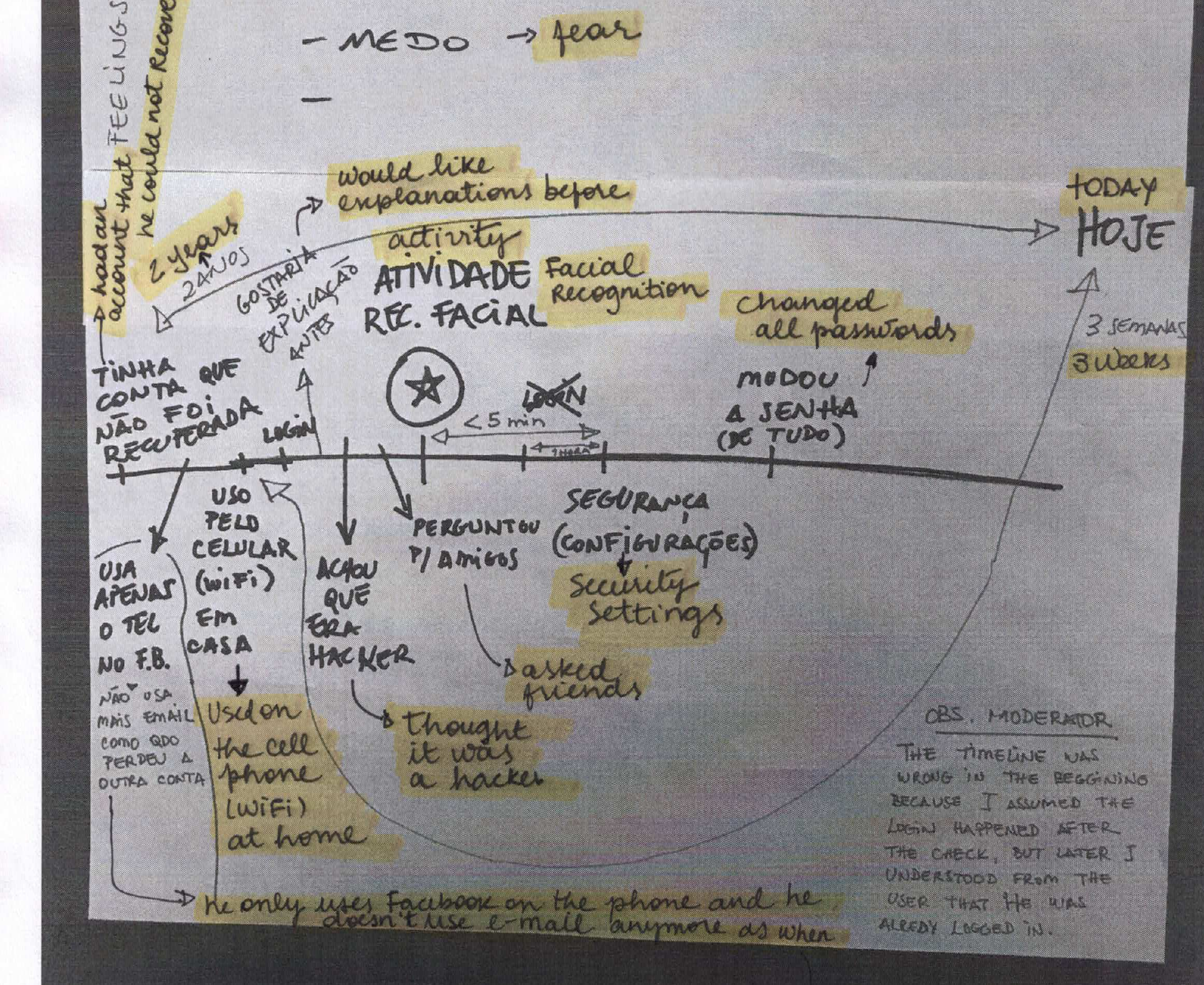}
\caption{\label{fig:br4} BR4 experienced the incident while using wifi at home. He thought that the incident was caused by someone trying to hack his account, but reached out to friends before completing the secondary login task to double check if they had ever experienced something similar. After completing the secondary authentication task, he checked his security settings for changes and then changed all of his passwords on both Facebook and his other social media accounts. He expressed that he was afraid when he thought his account was hacked, especially since 2 years ago he had lost access to a previous Facebook account. He also mentioned that he wished there had been information about what happened to his account in the notification, and that it would be cool if facial recognition was an option for a secondary authentication task.}
\vspace{-2ex}
\end{figure}

\subsection{Behavioral Response}
\label{sec:results:behavior}
Ultimately the mental model that participants generated about the incident informed their behavioral response: whether and what behavior to take to proactively secure their accounts and attempt to prevent future incidents. Participants who felt that an attacker had gained access to their account changed their password, while those who thought the notification was a false positive (e.g., because they were traveling) tended not to take any action. It is important to note that platforms are only able to predict that a suspicious login is legitimately an attack with some level of confidence: once informed, users similarly make their own prediction (or mental model), again with some confidence, about whether the incident is really an attack. Thus, lack of protective behavior is not necessarily a failing on the part of the participants. That said, the incidents about which participants in this study were notified were predicted to be authentic with the highest level of classifier confidence, while only a little over a third of users in our study, 24 participants, took a protective action. The majority (21 of 31) of the users who had a true-positive mental model took a protective action or checked for evidence of tampering, and a few (3 of 21) of the participants with weak mental models that centered around assuming the incident was a false positive, and who had not received notifications repeatedly, also took a protective action or checked their accounts.

Perhaps unsurprisingly, those participants with inaccurate mental models (e.g., those who thought that they had received the notification because of behavior the platform disapproved or that the incident was a ``random security check'') were less likely to take effective action (e.g., they would avoid liking posts) or any action at all. In fact, every participant who thought that they were selected randomly, or that everyone experienced the same incident on Facebook did not engage in a post-incident behavior. Those with weak mental models either chose multiple behaviors (a minority: 6 of 27) or did nothing at all (the remainder). For example, VN13 had a weak mental model consisting of multiple hypotheses for the incident. This led to confusion about what to do after regaining access to his account: ``I don't really know [which one happened]...and now what to do? I don't know'' (VN13). Finally, as aforementioned, 14 participants had experienced a similar Facebook security notification repeatedly, which led them to attribute the incident to something ``routine'' (DE5) that they did and feel annoyed about being notified. Consequently, none of these participants took a follow up action to further protect their account. Exemplifying this phenomena of fatigue, US3 says, ``it's just the same thing again, the only thing to do would be to stop [using private browsing]''.

\textbf{On-platform responses included changing passwords and settings, behaving ``better'', and checking accounts for tampering.} Twenty four of 67 respondents took an on-platform action in response to the incident and eight checked something (e.g., messages, settings) for evidence of tampering. This behavior was typically taken before any off-platform behavior. 

On Facebook, nine participants changed their password and five updated their privacy or security settings\footnote{Participants used the terms ``privacy settings'' and ``security settings'' interchangeably, so we do not distinguish in our analysis. In almost all cases, they were consulting the security settings page.}, while 10 changed their behavior on Facebook. When updating privacy / security settings, participants either added trusted contacts~\cite{trustedcontacts} to their account, set up two factor authentication, or set up SMS notifications of suspicious events. Of the latter, IN13 explains that setting up SMS notifications will keep their account more secure, because they will be able to complete the secondary authentication process immediately: ``now I put in my cellphone [number so] that I should receive alerts if someone tries logging into my Facebook account...so it won't be a surprise and I can kick them out right then.'' BR4 mentions changing his security settings to make sure his pictures were not publicly visible. He explains, ``if they're using those to protect my account, what good is it if everyone sees them?'' (Figure~\ref{fig:br4} contains the full timeline of BR4's incident experience). 

Those who changed their behavior on Facebook took a variety of different efforts to behave in ways they perceived as more safe or more approved by the platform. IN2 says ``I actually stopped adding strangers in my friend list and also stopped commenting on strangers' posts,'' because he associated these behaviors as ``unsafe'' and likely to have led to an attacker gaining access to his account (and thus, causing the notification). Similarly, VN4 says ``Now I would not click on something that is unclear, [like] sometimes I was tagged by my friends in some apps and auto comments, so [now], I will stay away from them. I will just click on the thing that I know, but for things that I don't know, I won't click them.'' Of the ten users who started avoiding certain behaviors, six had attributed the incident to having done something ``bad'', the other four had weak mental models and did multiple behaviors ``just in case'' (IN2). 
Participants also checked their Facebook accounts for tampering. Of those who did so, six checked their messages to see if any had been sent without their permission. For example, IN3 says, ``I checked the messages to see if there was anything [sent] deceiving other friends.'' 10 participants also checked their Facebook timelines for content that they had not posted or likes that they had not actioned; and five checked the Pages they ran on Facebook to see if anything had been posted or changed without their permission.

\textbf{Off-platform behaviors aim to reduce risk of encountering the incident again, but sometimes through security-compromising choices.} A smaller portion, 11 respondents, updated their off-platform security posture based on the particular incident we studied. BR4 (Figure~\ref{fig:br4}) not only adjusted his security settings, but ``also changed my password for other things too.'' While such adjustments may be positive, at times, users learned less advantageous lessons. IN9 mentions that he forgot his password and then, after repeatedly typing his old password, encountered the notification and secondary login process. To avoid this happening again he, ``changed [his] passwords on all my social media sites to be the same, so I don't forget''. Of the 11 respondents, it appears that three improved their security posture (one started using a password manager, two changed their passwords on other accounts to novel passwords), four made potentially less secure changes (saving passwords in browser, avoiding using VPN, using more similar or simpler passwords), and the remaining five made vague effort to be safer or more vigilant: e.g., ``I'm more careful on email [now] too'' (US5).

\textbf{Behavioral response is not always immediate.} For example, IN4 says ``The second time [I got a notification]...I was 
worried that I had been hacked...so now I have changed my password [and set my] privacy settings [to] very highly private. So, I am not worried now.'' While going through the notification once did not worry him, going through a second time did, and he subsequently decided to take action. Similarly, DE3 searched the Internet for information about the incident a few days after it occurred, deciding to change her password because ``the articles I saw seemed to say it was likely someone really was trying to get in and recommended changing the password.''

\textbf{Behavioral response was also informed through information seeking.} Seven participants in addition to DE3 sought out information about {\it behavior}: what to do after regaining access to their accounts. US3 says ``it wasn't clear if I was supposed to do something else or just go back to using the account...so I asked around to my friends about it''. Four of the eight participants who sought advice on behavior concluded that they should change their passwords, while two decided from the information they collected that they did not need to take any further action. The remaining participant was told ``my friend, he said, just be alert for the next few days, in case anything weird goes on in the account'' (IN12). Of these eight participants, all but one had what we characterize as weak mental models. They used outside information to bolster their understanding of what to do about what happened, even if they were still uncertain about the incident cause. 

\section{Incident-Response Information Seeking}
\label{sec:info}
As described in the prior section, participants sought out information about \textit{authenticity}, \textit{causation}, and/or {\it behavior} at different points in the response process. In this section we discuss {\it how} they sought information and their overarching motivation for doing so. We place these findings in the context of prior work on general security advice-seeking and consequently show that incident-response information is unique from general digital security information-seeking in the sources of support used, the urgency of seeking support, and motivations for seeking support.

\textbf{Sources of support and urgency of support.} A total of 23 participants sought out support during the incident response process. They used three support channels: 16 participants consulted informal sources (majority: family and friends), seven searched online, and five consulted the Facebook help pages (these five navigated directly to the Facebook help center, without first searching the Internet in general). BR4, for example, reached out to his friends. He thought that he knew what had happened to his account, ``I thought it had been hacked'' but wanted to see if ``any of my friends had experienced this, before doing something'' (Figure~\ref{fig:br4} shows his incident timeline). In contrast, prior work on support sources for digital security or Internet use shows that people often get advice about digital security through both informal (family, friends) and formal sources (librarians, workplace IT staff, paid support staff)~\cite{redmiles:2016, bakardjieva2005Internet,kiesler2000troubles,hargittai2003informed}. However, none of our participants sought information from a formal source and only three mentioned consulting a particular friend or family member because they were an expert. Additionally, again in contrast to prior work~\cite{redmiles:2016}, participants did not mention evaluating the information they received (either from people or from online sources) for quality. In the moment of incident response, users may be searching urgently for information, preoccupied with concern for their account rather than concern about the legitimacy of information they find: IN14 says, ``I tried to find it online. [I googled] ``Why is Facebook asking me to verify my comments, ask me to change my password and my privacy setting?'' I just wanted to find out what was wrong, fast!'' 

\textbf{Support facilitates camaraderie.} In addition to eliciting information that augmented participants technical understanding (authenticity, causation) and awareness of protective behaviors, for 14 of the 23 participants who sought out information seeking, this practice served to create a sense of \textit{camaraderie}. Illustrating this last point, IN4 says, ``I asked my friends...I came across 2 or 3 friends, they told me that their...accounts had been hacked...So, I thought, ok it's not just me.'' Similarly, US7 says ``I wanted to find out if this was normal, like something other people have had,'' and BR4 says ``Well I was at work and I asked people there whether they have received anything like that, they said no, so I was the only one that had experienced that.'' Pre-existing sense of camaraderie may also {\it reduce} motivation to seek information during incident response: for example, US4 says ``I thought [the secondary authentication process] happened sometimes with everyone so I didn't discuss [or seek out information].'' Participant emphasis on camaraderie may in part explain the prior success of using social influence to encourage security and privacy behavior~\cite{das2015role}. Finally, as camaraderie is not a motivation uncovered in prior work on general security information seeking, we hypothesize that camaraderie may be most relevant to in-the-moment, responsive security information-seeking such as that studied in this work.

\textbf{Participants who avoid seeking out information during or soon after an incident may consider the incident private or embarrassing.} Finally, those who consider account security (or other security and privacy) incidents private may avoid seeking out information. For example DE5 mentioned that ``I don't like to share personal things. I don't share anything personal like this [incident] with anyone, including my wife.'' They may also avoid doing so because they view seeking support as embarrassing: VN13 says, ``I didn't want to ask anyone because it was embarrassing, like I didn't know how to use Facebook. So I just hoped I was doing [the right thing].'' 11 of the 44 participants who did not seek out information reported avoiding doing so because this was a private incident (5 participants) or because looking for support would be embarrassing (6 participants). These sentiments of privacy and embarrassment around support have not been raised in prior security advice research. The majority of the remaining participants who did not seek out information had strong mental models, or had been repeatedly notified of similar incidents on the platform, and thus we hypothesize that they did not feel that they needed additional information.
%
%

\section{Influence of Cultural Context}
\label{sec:culture}
The processes of incident response we define in the prior sections was broadly consistent across participants  from five different countries with differing Internet and cultural contexts. However, we find initial trends that culture influences more latent components of users' experiences, which in turn drive different outcomes -- in terms of information seeking and protective behaviors -- from this shared process.

\textbf{Degree of Internet censorship} in the participant's country appears to influence their threat models: all six of the participants who described a government-related threat model were from more censored countries, India and Vietnam. These participants described ``the Spy'' threat model: that someone they knew would gain access to their account and share information with the government. Two of these participants also worried that someone could use account access to publicly disparage the government: ``if my account gets hacked and someone [says bad things about] prime minister Modi through my account, then it'll be a big problem for me. Maybe I'll not be able to stay in my country. That's a big problem'' (IN14).

\textbf{Collectivist cultural identity} appears to influence both participants' threat models and support sources. Participants from more collectivistic cultures (Vietnam, Brazil, and India) were more concerned about someone they knew gaining access to their account than an unknown ``hacker''. This finding aligns with results from prior work on Internet purchasing practices which showed significant differences in online risk perception between individualist and collectivist cultures~\cite{zheng2017does}. This emphasis on known attackers also led participants from Brazil and Vietnam to express concern about feelings of violation from account security incidents; these concerns were not raised by any participants from the US or Germany. BR13 explains, ``I would feel that someone was violating me. And I wouldn't know what to do because then I wouldn't be able to do anything to recover.'' Among Brazilian participants, this was, in fact, the second most-frequently mentioned concern about someone gaining access to their account.

Those from more collectivistic countries in our sample differed in their information sources: all but one of the participants from Brazil, Vietnam, and India who sought out information did so from a person. In contrast, those in the US were split regarding seeking out information from online sources vs. people, and only two participants in Germany mentioned seeking out information at all, one from an online search and one from a peer. Prior work has suggested that those who rely on people they know well (e.g., friends, family) have lower Internet skill than those who seek information online~\cite{courtois2014composition,helsper2017rich,redmiles:ccs}. However, our results preliminarily suggest that collectivist identities, in addition to potential variance in Internet skill due to recency of Internet adoption, may also explain variance in security information seeking channels.

\textbf{Differentiated platform use} also influences threat models among our participants, specifically regarding concerns about account access. For example, 10 of 17 participants in Vietnam used Facebook for business purposes compared to 2 of 11 participants in Germany, 1 of 9 in the US, 3 of 15 in India, and 3 of 15 in Brazil. Vietnamese participants describe threat models that include being concerned about financial consequences from someone accessing their Facebook account, far more than do participants from other countries. For example, VN6 explains: ``My page has a few tens [of] thousands of likes...if that admin's right was stolen, then I can't earn my living and they could use it for bad purpose.'' Similarly, fewer participants in Germany and India -- the two countries with fewest participants using Facebook Messenger -- mentioned being concerned about someone gaining access to their messages. 

Finally, prior cross-cultural work has focused quite broadly on the effects of culture or has considered factors related to a single cultural factor: \textbf{Internet penetration and skill}~\cite{sawaya2017self,kang2014privacy}. While it is possible that differentiated platform use is related to the recency of Internet adoption (or skill) of participants in a certain country, we observe no country-based usage patterns that indicate skill-based biases: for example, passive viewing of content on Facebook is equally prevalent in the US as in Brazil and Vietnam. Further, only six of 67 participants (from four of five countries) mentioned ``fear'' of not being able to complete the secondary authentication task, and no participants mentioned not engaging in a particular behavioral response due to concerns about skill / ability. 

The lack of skill-related influence that we observe is likely due to the fact that those in our sample were sufficiently skilled with online tasks to schedule an interview with us and complete a demographic survey, all online. While this limits our ability to comment on the relevance of skill in the response process, or the variance of skill by culture, it does allow us a unique opportunity to take an initial look at the influence of other, more sociological cultural factors (e.g., censorship, collectivism), which may otherwise be overshadowed by differences in skill. Further, it is encouraging that among our sample of users who were all able to complete the same online task (scheduling the interview): we do not observe country-level variance in incident response-relevant skills.

\section{Summary and Discussion}
\label{sec:discussion}
In this work, we inductively define a common process of account security incident response through an in-depth exploration of the experiences of users who had recently been notified of an account security incident. In sum, the notification process -- particularly the use of the secondary authentication task, which created a sense of platform-user partnership -- appears to be relatively effective both at alerting users to a threat and facilitating security action, when appropriate. 

While the incident notification process led to protective behavior for a third of all participants, and for the majority of those who thought the incident was a true positive, it was far less effective among participants who had weak mental models (e.g., those who were uncertain about what caused the incident). While the majority (51) of participants established plausible mental models for what had caused the incident, nearly half of those with plausible models established multiple plausible models -- it could have been a true positive, e.g., my brother trying to log in to my account, {\it or} a false positive, e.g., because I logged in from a new phone -- and were undecided between them. This lack of certainty regarding {\it why} the incident had happened and whether it was legitimate resulted in a lack of certainty about what to do next. While nine of these participants reached out for support from others in these cases, the notification itself was the primary source of information for the remaining participants who had weak mental models. 

Yet, notifications often lack key information -- particularly, information about the likelihood that this notification is informing the user of a legitimate threat. This lack of transparency can reduce notification efficacy. For example, users in our study who received repeated, frequent suspicious login notifications tended to increasingly believe that the notifications were false positives and that no protective action was needed. However, even if a ``more likely to be authentic'' event suddenly occurs -- like the incidents examined in this study -- participants may be given no indication that this notification or incident is different from prior, less risky incidents. Thus, they will likely not take any protective action. Beyond the example we present here, this phenomena has been echoed in work showing user fatigue toward SSL warning notifications~\cite{Sunshine:2009:CWE:1855768.1855793,felt2015improving}.

Recent work has shown that providing more transparency to users may help them make better decisions: a majority of users are able to make ``rational'' decisions about security when presented with concrete levels of hacking risk~\cite{redmiles2018dancing} and prior work in other fields shows that more information improves people's decision making~\cite{russo1974more,brewer1986choice,zarnoth1997social} and their trust in algorithmic predictions~\cite{bang2014does}. Thus, we argue that using UI indicators of classifier confidence in incident notifications may help to re-capture participant attention in high-risk situations and may help users make their own ``classification'' of the incident about which they have been notified.

In addition to improving classifier transparency, we hypothesize that incorporating user feedback into those security classifiers may further improve accuracy. Recent work in machine learning~\cite{grgic2018human,grgic2018beyond,kamar2012combining} has shown that integrating human inputs can improve classification results. Human intuition and knowledge -- in the context of suspicious logins, information about users' threat models, past experiences, and offline activities such as travel -- could thus be used as additional features for security classification. To this end, future work may explore how answers to short surveys after incident notifications -- for example, allowing users to rate their confidence that the incident was a true positive and provide detailed information about what they think happened -- can be fed back into security classifiers to improve incident identification accuracy. 

We hypothesize that creating this explicit feedback look between user and platform -- through post-incident user feedback and classifier transparency -- will enhance users' sense of partnership with the platform, which is already partially created through the secondary authentication process. We found that this sense of partnership increased users' sense of safety and engagement in the incident response process.

Finally, beyond the design of security classifiers and notifications, our results also have implications for secondary authentication mechanisms. The emphasis on known attackers among many of our participants not only informs future research, but also suggests potential vulnerabilities in existing secondary authentication methods: identifying pictures of friends may be difficult for a hacker who doesn't know you, but could be quite easy for someone who does. As shown in recent research on the privacy and security risks for domestic violence victims~\cite{matthews2017stories} account access by known attackers can be equally, or even more dangerous, than access by unknown attackers. As such, we urge focus on {\it known} attacker threat models in the ongoing development of user security mechanisms. 
%






%
\section*{Acknowledgements}
The author wishes to thank Mark Handel and Josh Rosenbaum, as well as Claudia Exeler and John Lyle, without whom this project would not have been possible. The author additionally wishes to thank Maximilian Golla, Ruogu Kang, Sean Kross, Michelle Mazurek and Franziska Roesner for their feedback and acknowledges support from the National Science Foundation Graduate Research Fellowship Program under Grant No. DGE 1322106 and a Facebook Fellowship.
\bibliographystyle{IEEEtran}

\end{document}